\documentstyle[twocolumn,psfig,aps]{revtex}
\tighten      
\begin{document}
\draft
\twocolumn[\hsize\textwidth\columnwidth\hsize\csname @twocolumnfalse\endcsname
\preprint{\vbox{Submitted to Physical Review C}}

\title{Nuclear dependence of the coherent\enspace \boldmath{$\!\!\eta$} 
       photoproduction reaction \\
       in a relativistic approach}
\author{L. J. Abu-Raddad$^{1,2}$, J. Piekarewicz$^{1}$, A. J. Sarty$^{2}$, and 
        M. Benmerrouche$^{3}$}
\address{${}^{1}$Supercomputer Computations Research Institute,\\ 
                 Florida State University, 
	         Tallahassee, FL 32306, USA}
\address{${}^{2}$Department of Physics, 
                 Florida State University, 
                 Tallahassee, FL 32306, USA}
\address{${}^{3}$Saskatchewan Accelerator Laboratory,
	         University of Saskatchewan, \\
		 Saskatoon, SK~S7N~5C6, Canada}
\date{\today}
\maketitle

 \begin{abstract}
  	We study the nuclear (or $A$) dependence of the coherent
  	$\eta$ photoproduction reaction in a relativistic impulse
  	approximation approach. We use a standard relativistic
  	parameterization of the elementary amplitude, based on a set
  	of four Lorentz- and gauge-invariant amplitudes, to calculate
  	the coherent production cross section from ${}^{4}$He, ${}^{12}$C, and
  	${}^{40}$Ca. In contrast to nonrelativistic treatments, our
  	approach maintains the full relativistic structure of the
  	process. The nuclear structure affects the process through the
  	ground-state tensor density. This density is sensitive to
  	relativistic effects and depends on $A$ in a different manner
  	than the vector density used in nonrelativistic
  	approaches. This peculiar dependence results in ${}^{4}$He
  	having a cross section significantly smaller than that of
  	${}^{12}$C---in contrast to existent nonrelativistic
  	calculations. Distortion effects are incorporated through an
  	$\eta$-nucleus optical potential that is computed in a simple
  	``$t\rho$'' approximation.
\end{abstract}


\vskip2pc]

\narrowtext


	The nuclear dependence of the coherent $\eta$ photoproduction
        process offers a unique opportunity to investigate medium
        modifications to the elementary $\gamma N \rightarrow \eta N$
        amplitude and might help distinguish between different
        theoretical models that provide an equally good description of
        the elementary process. In particular, the role played by the
        background is very significant in spin-isospin saturated
        nuclei where the dominant resonance $S_{11}$(1535) is
        suppressed. Furthermore, this reaction contributes 
        significantly to our understanding of nucleon-resonance
        formation and sheds some light on the propagation of these
        resonances through the nuclear medium. The $A$ dependence of
        this reaction is affected by the propagation of the produced
        $\eta$-meson through its interaction with the
        nucleus. Moreover, the coherent process is sensitive to the
        whole nuclear volume and, thus, depends on bulk properties of
        the nucleus. While nonrelativistic treatments suggest that
        nuclear-structure effects manifest themselves through the
        conserved vector (or baryon)
        density~\cite{bentan90,tryfik94,fix97}, our recent
        relativistic analysis suggests that, rather, it is the tensor
        density that affects the process~\cite{pisabe97}. This
	represents an important result, since the tensor density---a
	quantity as fundamental as the vector density---is not well
	determined by experiment.

	An early nonrelativistic study by Bennhold and Tanabe of the
	coherent $\eta$ photoproduction process predicted ${}^{4}$He
	to have the largest cross section of the three nuclei
	${}^{4}$He, ${}^{12}$C, and ${}^{40}$Ca~\cite{bentan90}.  A
	recent nonrelativistic study of this process seems to confirm
	this earlier prediction, although important quantitative
	differences do emerge~\cite{fix97}. In nonrelativistic
	treatments the coherent cross section is proportional to the
	square of the Fourier transform of the vector density. Thus,
	the particular $A$ dependence predicted by these calculations
	emerges from a competition between $A$, which tends to
	increase the cross section for larger nuclei, and the vector
	form-factor---which falls rapidly with $A$. This competition
	results in ${}^{4}$He having the largest cross
	section. Theoretical studies of this kind have motivated
	considerable experimental interest, which have culminated in
	an attempt to measure the coherent $\eta$ photoproduction cross
	section from ${}^{4}$He at the Mainz Microtron
	facility~\cite{ahrens93}.  Possibilities for extensions to
	higher energies and other nuclei exist, both at the Bonn ELSA
	facility and at TJNAF.

	In this report, as well as in our previous work on this
	subject~\cite{pisabe97}, we have used a relativistic approach
	to study this process. At no point in our calculation do we
	resort to a nonrelativistic reduction of the elementary
	amplitude or of the nuclear-structure model. Our results are
	in sharp contrast with the nonrelativistic predictions. Indeed, we
	find the coherent cross section from ${}^{12}$C as the
	largest, while that of ${}^{4}$He as the smallest one of the
	three. This is due to the relativistic
	character of the tensor---not the vector---density, which is
	the fundamental nuclear-structure quantity driving the
	reaction. It is this peculiar dependence of the tensor density
	with $A$ that is novel to our approach. Although the tensor
	density determines the qualitative behavior of the cross
	section with $A$, its quantitative behavior is determined by
	our choice of elementary amplitude. The elementary amplitude
	we have used in this work~\cite{benm92,bmz95} provides an
	excellent description of all available data on $\gamma p \rightarrow
	\eta p$ as well as the ones on $\gamma n \rightarrow
	\eta n$ as inferred from the very recent experiments on the deutron.


	The relativistic formalism for the coherent $\eta$
	photoproduction reaction, has been developed in our earlier
	work~\cite{pisabe97}. Thus, we will only reiterate here some
	of the main aspects of the formalism. The differential cross
	section in the center-of-momentum frame (c.m.) computed in a
	relativistic impulse-approximation approach is given by

\begin{equation}
   \left({d\sigma \over d\Omega}\right)_{\rm c.m.}=
   \; {\cal{K}}\; \; |F_{0}(s,t)|^{2} \;,
 \label{dsigmab}
\end{equation}
where
\begin{equation}
   {\cal{K}} \; \equiv \left({M_{\lower 1pt \hbox{$\scriptstyle T$}} 
    \over 4\pi W}\right)^{2} 
   \left({q_{\rm c.m.} \over k_{\rm c.m.}}\right)
   \left({1 \over 2}k_{\rm c.m.}^{2}q_{\rm c.m.}^{2}
   \sin^{2}\theta_{\rm c.m.}\right),
 \label{constant}
\end{equation}
        is a kinematical factor, and 
	$M_{\lower 1pt \hbox{$\scriptstyle T$}}$ is the mass of the
	target nucleus. Note that $W$, $\theta_{\rm c.m.}$, 
	$k_{\rm c.m.}$ and $q_{\rm c.m.}$ are the total energy, 
	scattering angle, photon and $\eta$-meson momenta in the
	c.m. frame, respectively. Hence,
	all dynamical information about the coherent process is
	contained in the single Lorentz-invariant form factor
        $F_{0}(s,t)$; this form-factor depends on the Mandelstam 
	variables $s$ and $t$.

	The Lorentz-invariant form-factor $F_{0}(s,t)$, is computed in
	a relativistic impulse-approximation approach. We use a
	standard, model-independent parameterization of the elementary
	$\gamma N \rightarrow \eta N$ amplitude. This elementary
	amplitude is given in terms of four Lorentz- and
	gauge-invariant amplitudes~\cite{bentan90,cgln57}. In
	nonrelativistic approaches it has been customary to evaluate
	this amplitude between on-shell Dirac spinors, thereby leading
	to the well-known Chew-Goldberger-Low-Nambu (CGLN) form of the
	elementary amplitude. In this work we do not resort to such a
	nonrelativistic reduction. Rather, we preserve the full
	relativistic content of the elementary amplitude and of the
	nuclear-structure model. In this way possible medium
	modification to the elementary process---that may arise from a
	different ratio of upper-to-lower components---can be
	examined.

	For closed-shell (spin-saturated) nuclei a significant
	simplification occurs, as the coherent process becomes
	sensitive to only one component of the elementary
	amplitude. In addition, all the nuclear-structure information
	is contained in the ground-state tensor
	density~\cite{serwal86}.  Thus, the Lorentz-invariant form
	factor---computed in a relativistic plane-wave impulse
	approximation (RPWIA) takes the following simple form:
\begin{equation}
  F_{0}^{\scriptscriptstyle PW}(s,t) = i A_{1}(\tilde{s},t) 
   {\rho_{\lower 3pt \hbox{$\scriptstyle T$}}(Q) / Q} \;.
  \label{fpwia}
\end{equation} 
	In this expression $\tilde{s}$ represents the effective (or
	optimal) value of the Mandelstam variable $s$ at which the
	elementary amplitude should be evaluated~\cite{cek87} and  
	$Q\equiv |{\bf k}_{\rm c.m.}-{\bf q}_{\rm c.m.}|\simeq
	\sqrt{-t}$. The ground-state tensor density is 
	defined by 
 \begin{eqnarray}
  \Big[
    \rho_{\lower 3pt \hbox{$\scriptstyle T$}}(r)\,\hat{r} \Big]^{i} &=& 
  \sum_{\alpha}^{\rm occ}
   \overline{{\cal U}}_{\alpha}({\bf x})\,
   \sigma^{{\scriptscriptstyle 0}i} \,
              {\cal U}_{\alpha}({\bf x}) \;,
 \label{rhotr}
 \end{eqnarray}
	where ${\cal U}_{\alpha}({\bf x})$ are the relativistic
	Dirac spinors. Note that in a relativistic plane-wave
	formalism the cross section is sensitive only to the 
	Fourier transform of the tensor density, i.e.,
\begin{equation}
    \rho_{\lower 3pt \hbox{$\scriptstyle T$}}(Q) =
     4\pi \int_{0}^{\infty} dr \, r^2
     j_{\lower 2pt \hbox{$\scriptstyle 1$}}(Qr)
    \rho_{\lower 3pt \hbox{$\scriptstyle T$}}(r) \;.
 \label{rhotq}
\end{equation}
	It is this tensor density that constitutes the fundamental 
	nuclear-structure quantity in this work. This is in
	contrast to nonrelativistic treatments that, instead,
	use the vector 
	density~\cite{bentan90,tryfik94,cek87,bofmir86,ndu91}. 
	The tensor density is a manifestation of the relativistic 
	character of this approach.

        We have computed the tensor density using a self-consistent,
	mean-field approximation to the Walecka 
	(or $\sigma$--$\omega$) model~\cite{serwal86}.  Even though the
	use of a mean-field approximation to describe a nucleus as
	small as ${}^{4}$He should be suspect, we feel justified in
	adopting this choice, as the coherent reaction is sensitive to
	only its bulk properties---which can be constrained by
	experiment.  Thus, in order to reproduce the experimental
	charge density of ${}^{4}$He, we have modified the mass of the
	$\sigma$ meson to $m_{\rm s}=564$~MeV---while maintaining
	constant the ratio of $g_{\rm s}^{2}/m_{\rm s}^{2}$.
	Note that we have used a standard set of parameters for the 
	Walecka model in our calculations of the ${}^{12}$C and
	${}^{40}$Ca nuclear structures: $g_{\rm s}^{2}=109.63$,
	$g_{\rm v}^{2}=190.43$,
	$m_{\rm s}=520$~MeV, and $m_{\rm v}=783$~MeV. 
	Finally, to achieve a more realistic picture of this process, 
	the plane-wave picture is modified by introducing interactions 
	(distortions) between the outgoing $\eta$ and the nucleus. 
	This is achieved by using an $\eta$-nucleus optical potential 
	of the $t\rho$ form~\cite{bentan90,pisabe97}. These
	distortions are sensitive to the ground-state vector density
	of the target nucleus.
	
\begin{figure}[h]
 \null
 \vskip1.0in
 \includegraphics{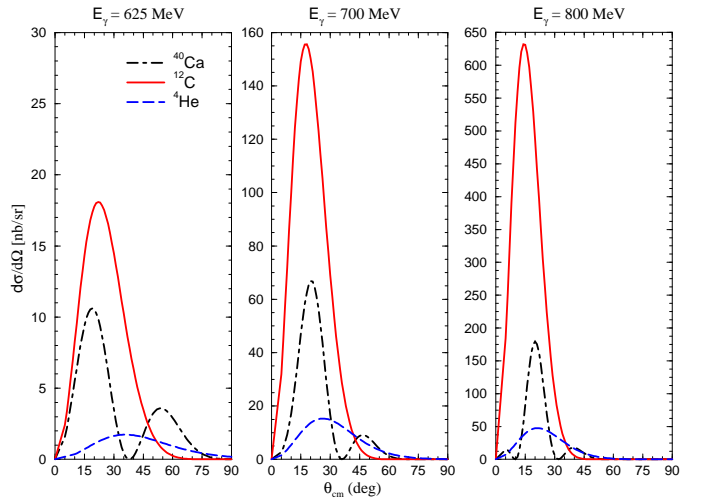}
 \vskip1.8in
\caption{The coherent $\eta$ photoproduction cross section as
         a function of A for photon laboratory energies of
 	 $E_{\gamma}=625$, $700$ and $800$~MeV, respectively.
	 All results were obtained using a RPWIA approach.}
\label{fig1}
\end{figure}
	The coherent $\eta$ photoproduction differential cross section from
	${}^{4}$He, ${}^{12}$C, and ${}^{40}$Ca is shown in Fig.~1 at
	photon laboratory energies of $625$, $700$, and $800$~MeV,
	respectively. Moreover, the total cross section as a function
	of the photon energy is shown in Fig.~2 for the same nuclei. No
	distortions have been included in these
	calculations. These results display significant relativistic
	corrections; there is a large enhancement of these cross
	sections relative to the nonrelativistic ones found in
	Refs.~\cite{bentan90,fix97}. This ``$M^{\star}$ effect''
	is a direct consequence of the enhancement of the lower 
	component of the Dirac spinor---which is determined 
	dynamically, rather than from a free-space relation.
	Moreover, there is an additional relativistic contribution 
	for open-shell nuclei, such as ${}^{12}$C; note that we are 
	treating ${}^{12}$C as a closed $p^{3/2}$ but open
	$p^{1/2}$ orbital. This can be most easily seen by assuming
	a free-space relation between the upper and lower components
	of the Dirac spinors. In this case the tensor density can be
	written in terms of the vector density as
\begin{equation}
  \rho_{\lower 3pt \hbox{$\scriptstyle T$}}(Q) =
  -{Q \over 2M_{N}} 
  \rho_{\lower 3pt \hbox{$\scriptstyle V$}}(Q) +
  \sum_{\alpha}^{\rm{occ}} {{\kappa +1} \over M_{N}} 
  \int_{0}^{\infty} dr {{g^{2}_{\alpha}(r)} \over r^2} j_1(Qr)\;,
 \label{trhotv}
\end{equation}
 	where $M_{N}$ is the free nucleon mass, $\kappa$ is the 
	generalized relativistic angular momentum, ${g_{\alpha}(r)}$ 
	is the upper component of the Dirac spinor, and 
	$j_1(Qr)$ is the Bessel function of order one. The second 
	term in the above expression is negligible for closed-shell
	nuclei; this term is proportional to the difference between 
	the square of the wave-functions of spin-orbit partners
	(such as $p^{3/2}$ and $p^{1/2}$ orbitals) which is very 
	small even in the Walecka model. Hence, for closed
	shell nuclei---and adopting a free-space relation---the
	tensor density becomes proportional to the vector density,
	as in the nonrelativistic approach. However, for open-shell 
	nuclei such as ${}^{12}$C, the second term in Eq.~\ref{trhotv} 
	is no longer negligible and leads to an additional enhancement 
	of the tensor density---above and beyond the one obtained from 
	the $M^{\star}$ effect.

\begin{figure}[h]
 \null
 \vskip1.0in
 \includegraphics{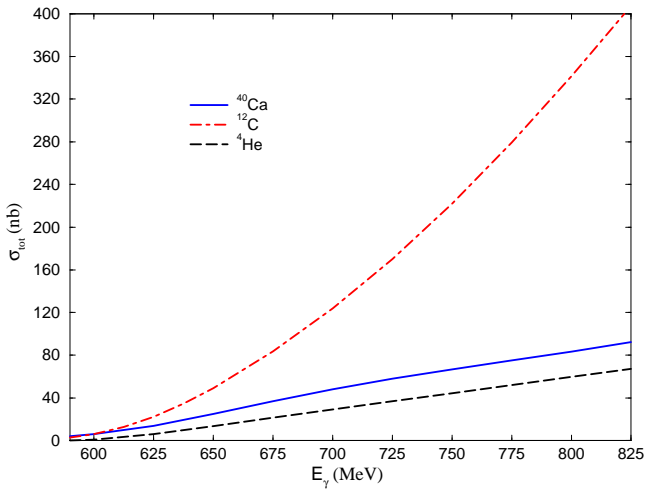}
 \vskip1.8in
\caption{The total coherent $\eta$ photoproduction cross section as
         a function of the incident photon laboratory energy from
         ${}^{40}$Ca, ${}^{12}$C, and ${}^{4}$He.
	 All results were obtained using a RPWIA approach.}
\label{fig2}
\end{figure}
 	In Fig.~3 the cross section from the same three nuclei is
	displayed with distortions added to the emitted $\eta$ by
	using a relativistic distorted-wave-impulse approximation
	(RDWIA). Since at low energy (e.g., $E_{\gamma} = 625$~MeV)
	the real part of the optical potential is attractive, its
	competition with the (absorptive) imaginary part produces a
	distorted-wave cross section that differs little from its
	plane-wave value. However, at higher energies the real part
	becomes repulsive, leading to a substantial reduction in the
	value of the cross section. For a small nucleus such
	${}^{4}$He the effect of distortions are less pronounced than
	in ${}^{12}$C and in ${}^{40}$Ca. This is consistent with the
	standard picture that emerges from nonrelativistic
	calculations~\cite{bentan90}.

\begin{figure}[t]
 \null
 \vskip1.0in
 \includegraphics{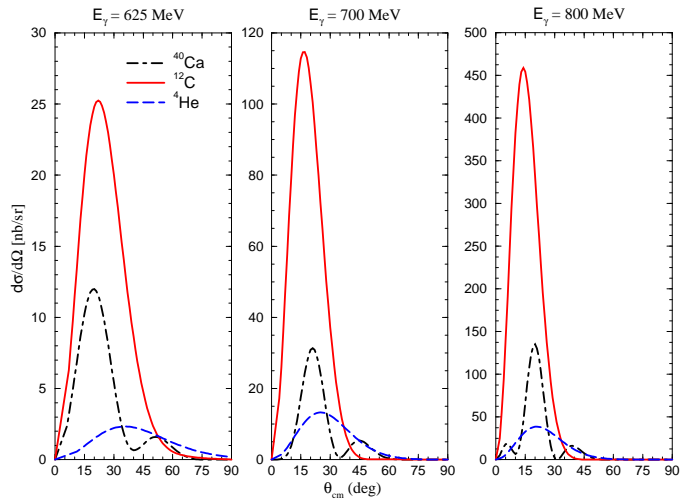}
 \vskip1.8in
\caption{The coherent $\eta$ photoproduction cross section as
         a function of A for photon laboratory energies of
 	 $E_{\gamma}=625$, $700$ and $800$~MeV, respectively.
	 All results were obtained using a RDWIA approach.}
\label{fig3}
\end{figure}	
 	Our relativistic results differ significantly from those
	obtained in nonrelativistic calculations. Indeed, Bennhold and
	Tanabe~\cite{bentan90} have predicted that ${}^{4}$He would
	have the largest cross section of the three nuclei, due to
	its largest charge form factor. This, we believe, might have
	been an important reason to select ${}^{4}$He for the first
	experimental measurement of the coherent process. However,
	this finding is at odds with our relativistic results, which 
	instead show ${}^{4}$He to have the smallest cross section 
	as can be seen in Figs.~1 and 2. There are two main reasons 
	for this differences. First, in relativistic calculations 
	the ratio of upper-to-lower components is determined dynamically 
	in the Walecka model, rather than from its free-space relation. 
	The Walecka model is characterized by strong scalar and vector 
	potentials that generate an enhancement in the lower component 
	of the wavefunction---and a corresponding enhancement in the tensor 
	density. Second, the elementary $\eta N$ interaction used in this
	work~\cite{benm92,bmz95} is different from the one used 
	by Bennhold and Tanabe~\cite{bentan90}, in particular the non
	resonant contributions were not considered in the
	latter. Although both models
	seems to give an adequate description of the elementary process,
	important differences emerge in the calculation of the 
	coherent reaction. This is primarily due to the fact that
	the coherent process from spin-saturated nuclei becomes
	insensitive to the dominant 
	$S_{11}$(1535) intermediate-resonance contribution, and
	therefore quite sensitive to the details of other resonant and
	non-resonant background contributions such as the
	$D_{13}(1520)$ and vector mesons. Note that our calculations
	for ${}^{4}$He are similar to the nonrelativistic ones reported
	recently by Fix and Arenh\"ovel's~\cite{fix97}. However, this 
	agreement seems to be fortuitous, since neither their 
	nuclear-structure model nor their elementary amplitude 
	are similar to ours; their coherent process is dominated 
	by $\omega$-meson exchange, while ours contains, in addition, 
	a significant contribution from the $D_{13}$(1520) resonance.


	In conclusion, the main goal of our present work was to
	elucidate the $A$ 
	dependence of the coherent $\eta$ photoproduction cross
	section in a relativistic impulse-approximation approach.
	We found the cross section sensitive to two 
	nuclear-structure quantities: {\it i)} the ground-state
	vector density and {\it ii)} the ground-state tensor
	density. The tensor density is as fundamental as the vector 
	density used in the nonrelativistic treatment, although it 
	is not as well constrained by experiment. 

	We have found important discrepancies vis-a-vis
	nonrelativistic results. Part of these discrepancies stem from
	the fact that we have used a fully relativistic
	approach---with no resort to a nonrelativistic
	reduction. Moreover, the elementary amplitude used in our
	model is different from the ones used in other theoretical
	calculations~\cite{bentan90,fix97}. Our relativistic approach
	suggests the use of the tensor density as the fundamental
	nuclear-structure quantity driving the reaction. Although our
	results are also sensitive to the vector density (through
	distortion effects) for a small nucleus such as ${}^{4}$He,
	or at low-energies (where the real part of the optical
	potential is attractive) distortion effects become small and
	the relativistic cross section becomes dominated by the the
	tensor density. The tensor density, as opposed to the vector
	density, is sensitive to the relativistic corrections arising
	in the nuclear medium. The use of the tensor density
	represents one of the central results of our treatment.

	Many challenges remain. First, one should try to study
	possible violations to the impulse-approximation picture.
	Second, there are off-shell ambiguities in the elementary
	amplitude. The form of the elementary amplitude used here is
	standard but not unique. There are many other choices which
	are equivalent on-shell, but can give vastly different results
	off-shell~\cite{cgln57,ndu91}. Although there are some
	attempts to deal with this issue~\cite{fix97}, a detailed 
	microscopic model is needed to take the amplitude off-shell.
 	Finally, as the coupling to the intermediate $S_{11}$(1535)
	resonance dominates
	the elementary $\gamma N \rightarrow \eta N$---but not the
	coherent process---the coupling to additional resonances
	is poorly determined. Indeed, while 
	Fix and Arenh\"ovel~\cite{fix97} suggest a negligible $D_{13}$(1520) 
	contribution to the coherent process, our elementary model 
	predicts a significant one.

	Undoubtedly, there is still a lot of work to be done both
	experimentally and theoretically. We hope that with the 
	advent of new powerful and sophisticated facilities, such
	as TJNAF and MAMI, the validity of the different theoretical
	models could be tested. This could help us elucidate the
	underlying mechanism behind the coherent $\eta$ photoproduction
	process.

This work was supported in part by the U.S. Department of Energy
under Contracts Nos. DE-FC05-85ER250000 (JP), DE-FG05-92ER40750 (JP), 
by the U.S. National Science Foundation (AJS), and by the Natural 
Sciences and Engineering Research Council of Canada (MB).

\medskip

\vfil

\end{document}